# Level generation for rhythm VR games

Bachelor's Thesis

**Mariia Rizhko**

Supervised by Andrii Boichuk

Lviv, Spring 2022

# ABSTRACT

Ragnarock is a virtual reality (VR) rhythm game in which you play a Viking captain competing in a longship race. With two hammers, the task is to crush the incoming runes in sync with epic Viking music. The runes are defined by a beat map which the player can manually create. The creation of beat maps takes hours.

This work aims to automate the process of beat map creation, also known as the task of learning to choreograph. The assignment is broken down into three parts: determining the timing of the beats (action placement), determining where in space the runes connected with the chosen beats should be placed (action selection) and web-application creation.

For the first task of action placement, extraction of predominant local pulse (PLP) information from music recordings is used. This approach allows to learn where and how many beats are supposed to be placed.

For the second task of action selection, Recurrent Neural Networks (RNN)[10][11] are used, specifically Gated recurrent unit (GRU)[12][13] to learn sequences of beats, and their patterns to be able to recreate those rules and receive completely new levels.

Then the last task was to build a solution for non-technical players, the task was to combine the results of the first and the second parts into a web application for easy use. For this task the frontend was built using JavaScript and React and the backend - python and FastAPI.

# CONTENT

# LIST OF ABBREVIATIONS

VR - Virtual Reality

RNN - Recurrent Neural Networks

CNN - Convolutional Neural Networks

GRU - Gated recurrent unit

MFCC - Mel-frequency cepstral coefficients

DFF - Discrete Fourier Transform

DCT - Discrete Cosine Transform

MLSTM - Multiplicative LSTM

PLP - Predominant local pulse

JSON - JavaScript Object Notation

# 1 INTRODUCTION

Ragnarock [1] is a rhythm-based VR game released at the end of 2021 and becoming more and more popular now. It has about 20 build-in levels and the possibility to buy additional seven songs. There is nothing to say that players are mastering those levels pretty quickly and want to play new songs. Though the game allows custom maps, the manual creation of those can take from several hours up to a day.

The game's goal is to use two hammers in the rhythm of the music and crush runes when they reach the drums. A snapshot showing the gameplay of Ragnarock is shown in Figure 1.

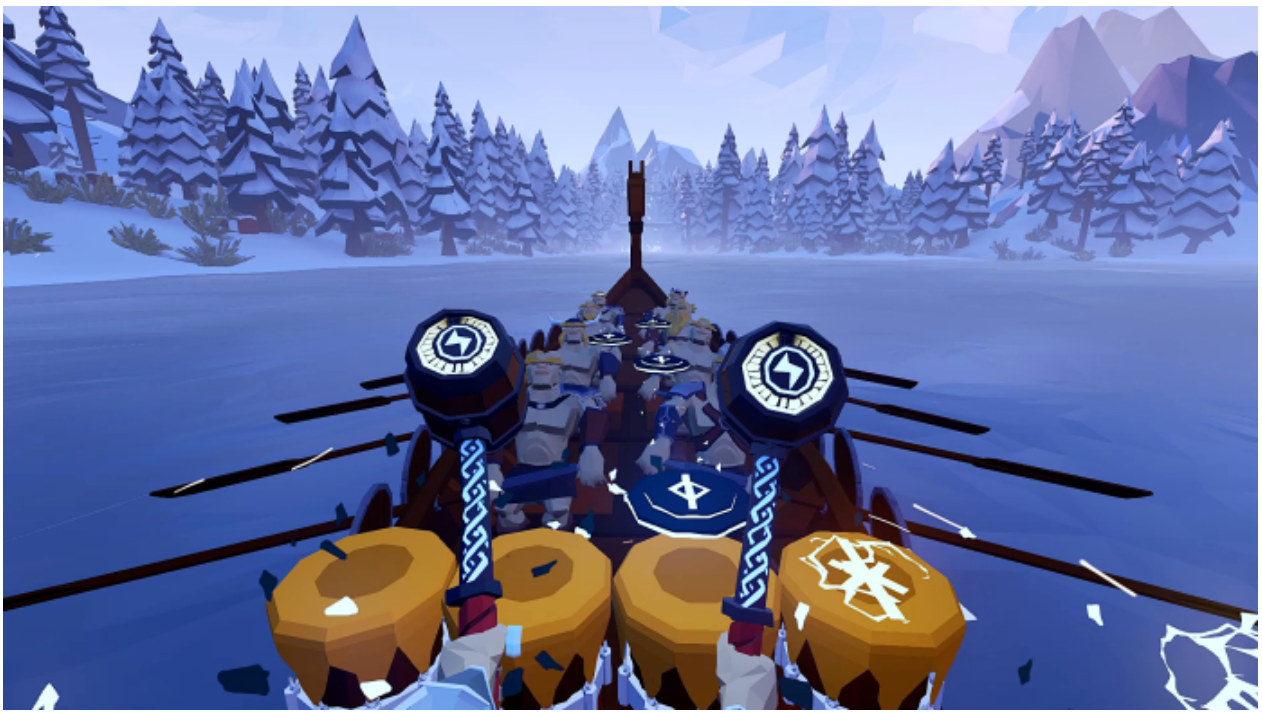

Figure 1: Gameplay of Ragnarock

This game is not the first game of this type. Before it, there was a Beat Saber [2]. It was the most popular VR game of 2019 [7]. It depicts the user slicing blocks representing musical beats with a pair of contrasting-colored sabers in a surrealistic neon atmosphere.

The player controls a pair of swords with VR motion controllers, colored red and blue for left and right, respectively. Each song has a stream of incoming blocks. Each block appears at a specific time according to the beats of the music. It can be placed in one of 12 spots of a 4x3 grid. Each is marked with an arrow pointing in one of eight different directions. Players can also hit blocks with dots in any direction. When a saber cuts a block appropriately, it is destroyed, and a score is given based on the length.

The goal of this game is similar to Ragnarock, but instead of crushing runes, you slice cubes. You can see a snapshot showing the gameplay of Beat Saber in Figure 2.

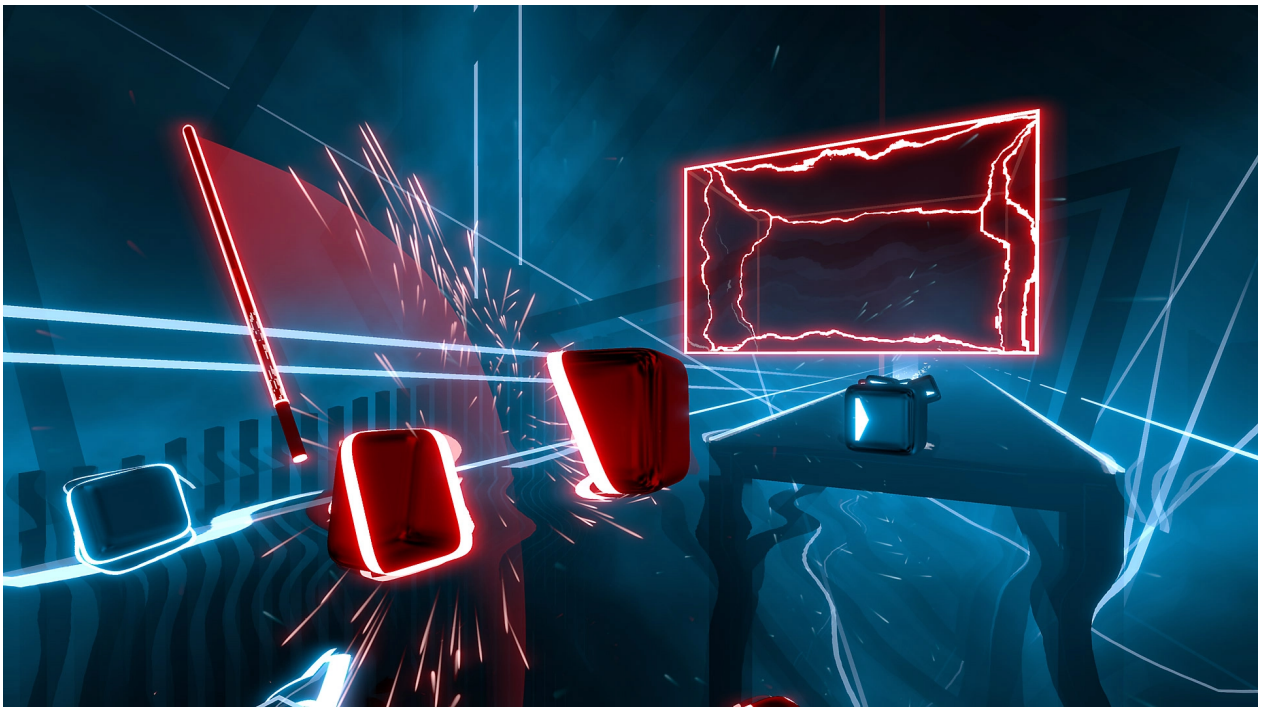

Figure 2: Gameplay of Beat Saber

Another game with similar gameplay but way older is Dance Dance Revolution. It is the pioneering rhythm and dance genre series in video games, first released in Japan in 1998 and later released in North America and Europe in 1999. To musical and visual cues, players stand on a "dancing platform" or stage and hit colorful arrows spread out in a cross with their feet. Players are scored on how well they time their dance to the patterns offered to them, and if they pass, they are given more music to play. So here, you are supposed to step on arrows instead of crushing runes or slicing cubes. You can see a snapshot showing the gameplay of Dance Dance Revolution in Figure 3.

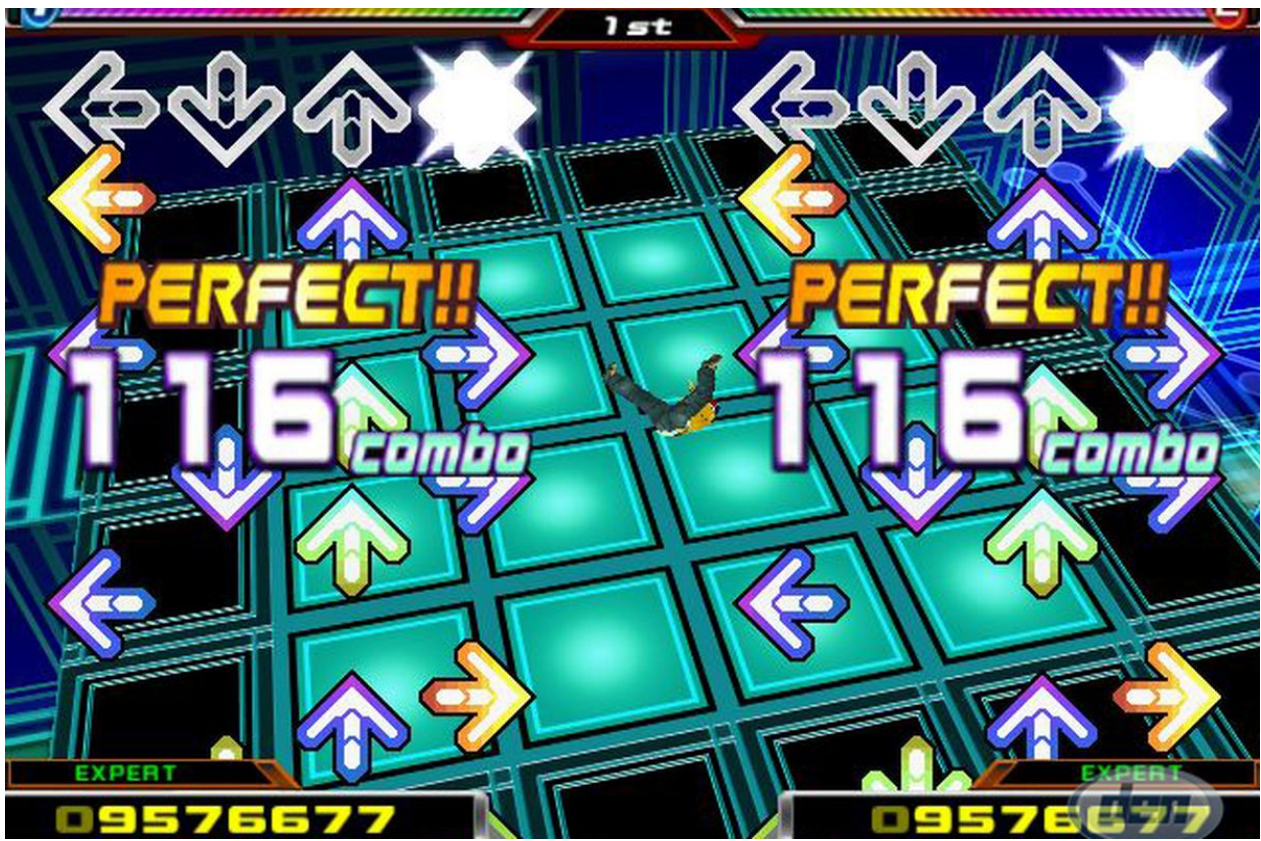

Figure 3: Gameplay of Dance Dance Revolution

Although all these games have different mechanics and elements: in Dance Dance Revolution, it's making steps on arrows; in Beat Saber, it's slicing cubes; and in Ragnarock, it's crushing runes, the main idea behind levels are the same. The specific element occurs at a particular time. The first time this task was introduced in Dance Dance Convolution [3] as a task of learning to choreograph.

The goal of this work is to learn to choreograph for the game Ragnarock. The task is separated into three subtasks. The first one is action placement, which has to generate the sequence of timestamps when to put the element. The second one is action selection, which is to figure out which exact elements to put. And the last one is web application development.

The main contribution of this work is solving the problem of learning to choreograph for the game Ragnarock and developing a web application for non-technical users.

# 2 RELATED WORKS

## 2.1 Dance Dance Revolution

Dance Dance Convolution [3] was the first work that tried to work with the problem of level generation for rhythm-based games. They introduced it as a task of learning to choreograph, designed and evaluated several deep learning methods, and introduced standardized datasets and reproducible evaluation methodology.

They decomposed the task into two subtasks: deciding when to place steps and which steps to select. For the step placement task, they combined recurrent and convolutional neural networks to ingest spectrograms of low-level audio features to predict actions conditioned on chart difficulty. They presented a conditional LSTM[15] generative model for step selection that substantially outperformed n-gram and fixed-window approaches.

Their pipeline was the following:

1. Represent audio as features
2. Estimate probabilities that step lies within the frame by feeding the representation from step 1 into a step placement algorithm
3. Feed the sequence of probabilities from step 2 into a peak-picking algorithm to identify the precise timestamps at which to place steps
4. Use a step selection algorithm to choose which steps to place at each time into given a sequence of timestamps

## 2.2 Beat Saber

DeepSaber: Deep Learning for high dimensional choreography [4] was based on Dance Dance Convolution work. Multiple Beat Saber map generation approaches were described, implemented, and evaluated. They demonstrated the similarity between natural language and beat maps and used it to advance beat map generation technology. They created an action analogy dataset and used it to show how high-quality action embeddings are. They used action embeddings to develop a new training metric and a novelty-based metric to compare entirely generated beat maps to maps created by humans.

For audio processing, they used Mel-frequency cepstral coefficients (MFCC)[14] features, applied pre-emphasis, took the Discrete Fourier Transform (DFF), applied 40 triangular Mel-filterbanks, applied log, computed Discrete Cosine Transform (DCT), took the first 13 cepstral coefficients, replace them by a log of frame energy for to account for volume and finally shifted the timestamps.

The number of dimensions in inputs in Beat Saber varies greatly. Each input's optimal number of parameters for expressing abstract representation should be different. So they used a Multi LSTM (MLSTM) architecture to find a good architecture that addressed the difference in input dimensionality. Because the connections between LSTM blocks are random, most architectures produced poor results. They solved this problem by using a much larger number of epochs (110) in the Hyperband algorithm.

# 3 MATERIALS AND METHODS

## 3.1 Building dataset

The game Ragnarock allows manually created maps, so the community created hundreds of maps and placed them on web site [5]. The screenshot of this site is presented in Figure 4.

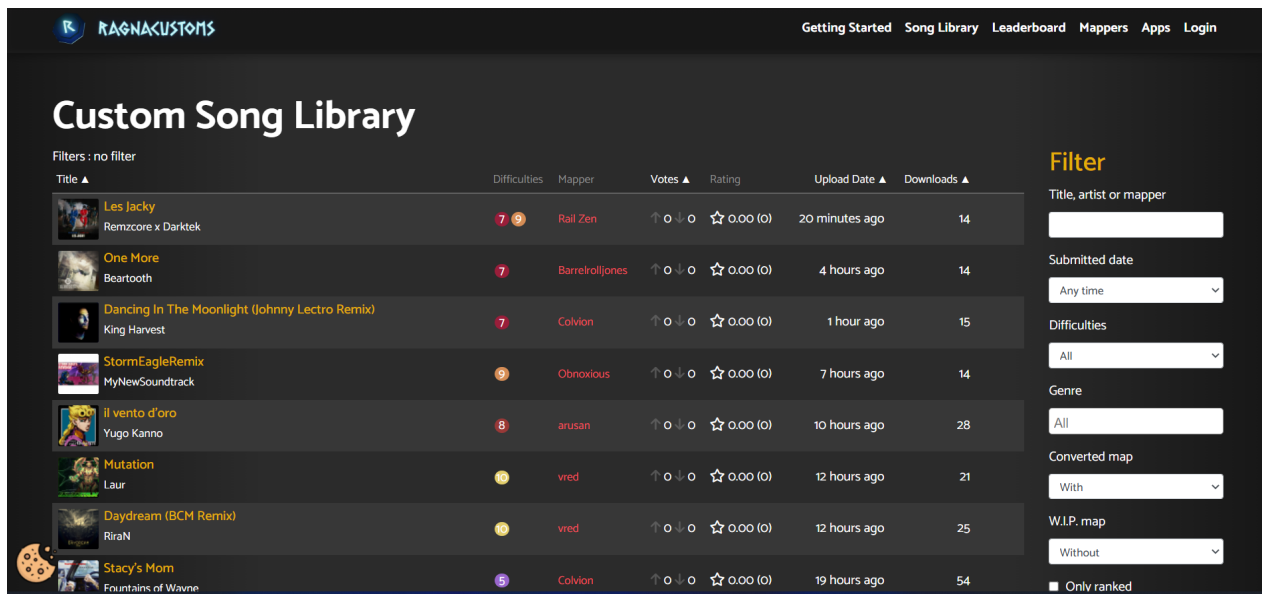

Figure 4: The site of manually created levels

To create a dataset, levels from this site were parsed. In total, there were 831 songs. For each song, the following data was parsed: song id, song name, song author, a song file name, beat per minute (bpm), whether easy, normal, hard, expert, or expert plus difficulty exists.

In total, there were different amounts of maps with each difficulty:

- Easy: 572
- Normal: 130
- Hard: 95
- Expert: 147
- Expert Plus: 186.

Each game level is presented as a folder that users can put inside the game and play a song. In this folder there are next files:

- Info.dat: obligatory file that contains general information about the level
- Song file: obligatory audio file in .ogg format
- Song preview file: obligatory short version of the song file, also in .ogg format
- Cover image file: obligatory picture for level preview
- Easy.dat: optional file for easy difficulty
- Normal.dat: optional file for normal difficulty
- Hard.dat: optional file for hard difficulty
- Expert.dat: optional file for expert difficulty
- ExpertPlus.dat: optional file for expert plus difficulty

Info.dat file has general info about the level such as:

- songName
- songSubName
- songAuthorName
- levelAuthorName
- beatsPerMinute
- songFilename
- coverImageFilename
- difficultyBeatmapSets
- etc

difficultyBeatmapSets is a list of difficultyBeatmaps. Each one has the following info:

- difficulty

- difficultyRank
- beatmapFilename
- noteJumpMovementSpeed
- noteJumpStartBeatOffset

beatmapFilename is the name of the level difficulty map (such as Easy.dat, Normal.dat, Hard.dat, etc). This difficulty map is a dictionary, one of the keys is _notes. _notes is a list of beats each of them has:

- _time
- _lineIndex
- _lineLayer
- _type
- _cutDirection

An example of such a file is shown in Figure 5.

```
▼ _notes: [] 679 items
  ▼ 0:
      _time: 4
      _lineIndex: 0
      _lineLayer: 1
      _type: 0
      _cutDirection: 1
  ▼ 1:
      _time: 4
      _lineIndex: 1
      _lineLayer: 1
      _type: 0
      _cutDirection: 1
  ▼ 2:
      _time: 4
      _lineIndex: 2
      _lineLayer: 1
      _type: 0
      _cutDirection: 1
  ▼ 3:
      _time: 4.5
      _lineIndex: 1
      _lineLayer: 1
      _type: 0
      _cutDirection: 1
```

Figure 5: Example of beatmapFilename

The first task of action placement is the task of setting _time parameter for each beat. The second task of action selection is setting _lineIndex, _lineLayer, _type, and _cutDirection for each beat. Moreover, the last task of app creation is setting up all other parameters based on user choice, saving files in one folder, and sending it to a user.

## 3.2 Data processing

As described in section 3.1, the files are presented in the form of dictionaries, which means it requires additional preprocessing before using in any machine learning algorithm.

Each beatmapFilename was parsed into a list of elements. Each element represents a unique id depending on beat's _lineIndex, _lineLayer, _type, and _cutDirection. The unique id is built as a flat representation of three by four greed. This is made in several steps:

1. Select the timestamp with at least one element presented at it
2. Assign the default value [['n9', 'n9', 'n9', 'n9'], ['n9', 'n9', 'n9', 'n9'], ['n9', 'n9', 'n9', 'n9']] to the element
3. Change the value in grid on position (_lineLayer, _linIndex) to be _type + _cutDirection, where + is an operation of string concatenation
4. If there are other elements at the same timestamp, do step 3 for them
5. Flatten an array into a string

The result of preprocessing is shown in Figure 6.

```
▼ root: [] 531074 items
    0:  "n9n9n9n901n9n901n9n9n9n9"
    1:  "n9n9n9n9n901n9n9n9n9n9"
    2:  "n9n9n9n9n901n9n9n9n9n9"
    3:  "n9n9n9n9n9n901n9n9n9n9"
    4:  "n9n9n9n901n9n9n9n9n9n9"
    5:  "n9n9n9n9n901n9n9n9n9n9"
    6:  "n9n9n9n90101n9n9n9n9n9"
    7:  "n9n9n901n9n901n9n9n9n9"
    8:  "n9n9n9n9n901n9n9n9n9n9"
    9:  "n9n9n9n901n9n9n9n9n9n9"
    10: "n9n9n9n901n9n9n9n9n9n9"
    11: "n9n9n9n9n9n901n9n9n9n9"
    12: "n9n9n9n901n9n9n9n9n9n9"
    13: "n9n9n9n90101n9n9n9n9n9"
    14: "n9n9n9n9n9n901n9n9n9n9"
    15: "n9n9n901n9n9n9n9n9n9n9"
    16: "n9n9n901n9n9n9n9n9n9n9"
    17: "n9n9n9n9n901n9n9n9n9n9"
    18: "n9n9n9n901n9n9n9n9n9n9"
    19: "n9n9n90101n9n9n9n9n9n9"
    20: "n9n9n901n9n901n9n9n9n9"
    21: "n9n9n901n9n9n9n9n9n9n9"
    22: "n9n9n9n9n901n9n9n9n9n9"
    23: "n9n9n9n9n901n9n9n9n9n9"
    24: "n9n9n901n9n9n9n9n9n9n9"
    25: "n9n9n9n901n9n9n9n9n9n9"
    26: "n9n9n90101n9n9n9n9n9n9"
    27: "n9n9n90101n9n9n9n9n9n9"
    28: "n9n9n901n9n901n9n9n9n9"
```

Figure 6: Example of preprocessed beatmapFilename

# 4 EXPERIMENTS

## 4.1 Action placement

The task is to find timestamps where to place beats with raw audio, as displayed in Figure 7.

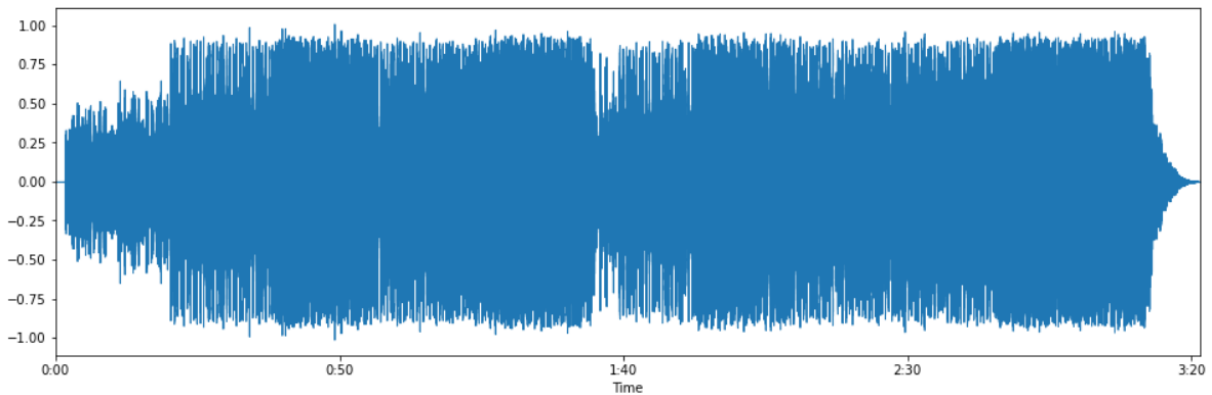

Figure 7: Raw audio file

As Ragnarock is a rhythm-based game, most beats' location depends on the rhythm of the music track. So theoretically, there should be a correlation between the dynamic pulse of a song and beats location. The Predominant local pulse (PLP)[6][8][9] estimation is used. The PLP method analyzes the onset strength envelope in the frequency domain to find a locally stable tempo for each frame. These local periodicities synthesize local half-waves, which are combined such that peaks coincide with rhythmically salient frames (e.g., onset events on a musical time grid). The local maxima of the pulse curve can be taken as estimated beat positions.

The example of applying PLP to an audio sample of 1000 frames and then local maxima to PLP is shown in Figure 8.

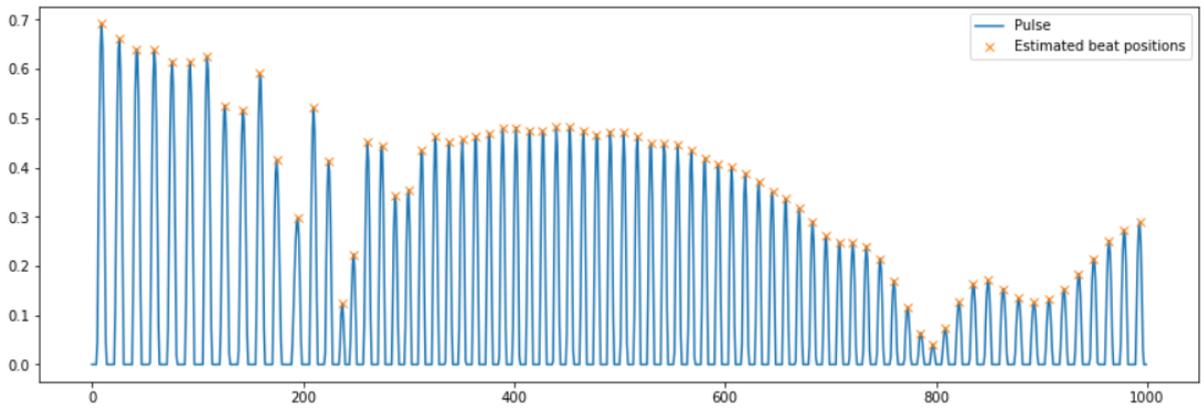

Figure 8: PLP and estimated beat positions

The blue line represents the change in local pulse through frames. Orange crosses - estimated beat positions found by applying local maxima algorithm to PLP.

Estimated beat positions are the ideal solution, but the actual data differs. In figure 9, actual beat positions over PLP are shown.

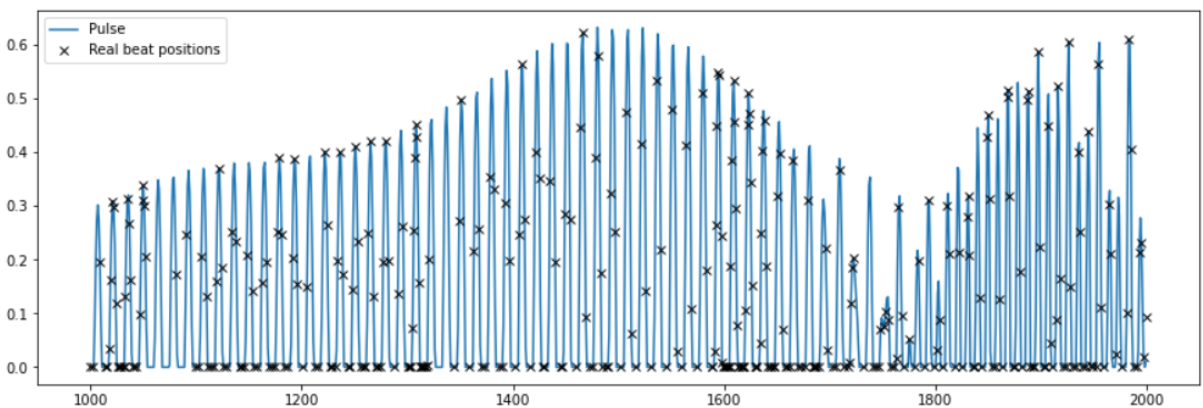

Figure 9: PLP and actual beat positions

The data generally falls for PLP changes, but not ideally. The 1-D smoothing spline with a degree of the smoothing spline 3 fits estimated and actual data to prove the correlation between estimated and actual beat positions. The splines are displayed in Figure 10.

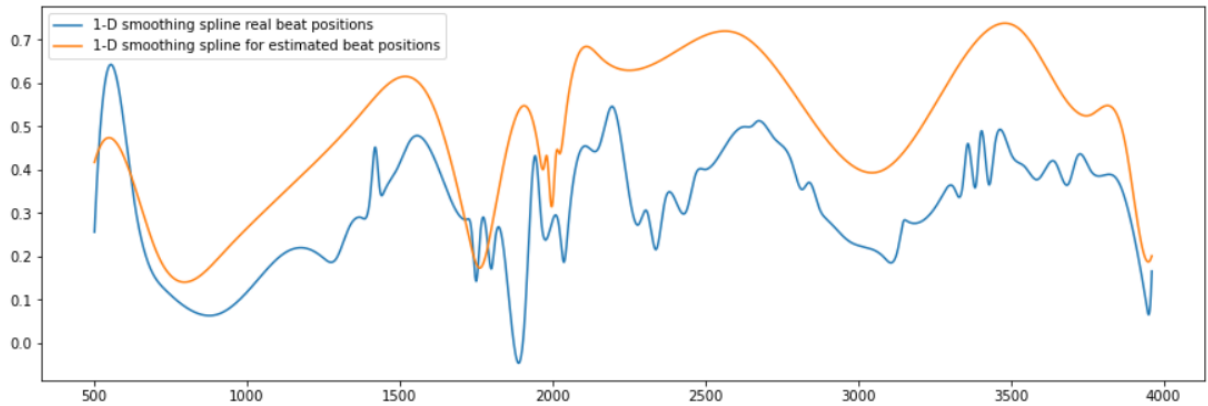

Figure 10: Splines fit actual and estimated beat positions.

From the figure above it is possible to assume there is direct positive correlation between estimated and real beat positions. A correlation coefficient between these two splines equals 0.75, which means PLP and then local maxima search are good algorithms for the action placement task.

Another question that needs to be answered is the number of beats to place. The higher level in the game, the more beats should appear. It is possible to change the amount of peaks selected during the local maxima search by changing the distance parameter. Required minimal horizontal distance must be bigger or equal to 1 in samples between neighboring peaks. Smaller peaks are removed first until the condition is fulfilled for all remaining peaks. The bigger distance, the smaller amount of peaks selected. An example of reducing beats amount is shown in Figure 11.

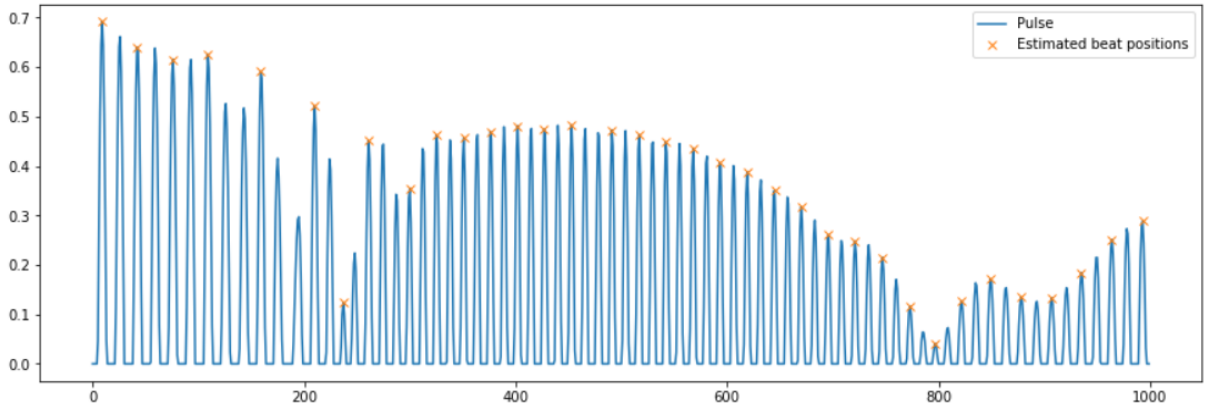

Figure 11: PLP and reduced amount of estimated beat positions

To determine the optimal distance for different levels, ten samples for each difficulty were selected with their correspondent song duration, bpm from info, bpm from librosa library, and a number of beats. Selected data is shown in Figures 12-16.

| | diff | duration | bpm_info | bpm_librosa | num_el | | diff | duration | bpm_info | bpm_librosa | num_el |
|---|---|---|---|---|---|---|---|---|---|---|---|
| 0 | Easy | 190.827256 | 102.0 | 103.359375 | 505 | 10 | Normal | 233.887392 | 91.0 | 92.285156 | 781 |
| 1 | Easy | 124.226757 | 120.0 | 112.347147 | 362 | 11 | Normal | 254.189841 | 135.0 | 135.999178 | 1463 |
| 2 | Easy | 258.647075 | 320.0 | 117.453835 | 667 | 12 | Normal | 230.365170 | 88.0 | 89.102909 | 671 |
| 3 | Easy | 198.089478 | 130.0 | 129.199219 | 656 | 13 | Normal | 269.373333 | 178.0 | 135.999178 | 1929 |
| 4 | Easy | 237.621451 | 82.0 | 161.499023 | 823 | 14 | Normal | 271.037732 | 213.0 | 107.666016 | 1142 |
| 5 | Easy | 189.141088 | 105.0 | 107.666016 | 628 | 15 | Normal | 190.721451 | 174.0 | 117.453835 | 627 |
| 6 | Easy | 516.585986 | 96.0 | 129.199219 | 2053 | 16 | Normal | 118.767528 | 148.0 | 143.554688 | 284 |
| 7 | Easy | 273.501270 | 98.0 | 99.384014 | 608 | 17 | Normal | 136.080590 | 126.0 | 129.199219 | 566 |
| 8 | Easy | 269.545034 | 87.0 | 172.265625 | 764 | 18 | Normal | 150.793016 | 130.0 | 129.199219 | 717 |
| 9 | Easy | 243.577370 | 140.0 | 143.554688 | 897 | 19 | Normal | 262.452245 | 160.0 | 161.499023 | 458 |

Figures 12-13: Selected data for Easy and Normal difficulties

| | diff | duration | bpm_info | bpm_librosa | num_el |
|---|---|---|---|---|---|
| 20 | Hard | 132.021905 | 150.000000 | 151.999081 | 412 |
| 21 | Hard | 168.741950 | 155.000000 | 103.359375 | 655 |
| 22 | Hard | 234.979955 | 105.000000 | 103.359375 | 1014 |
| 23 | Hard | 164.633651 | 174.000000 | 172.265625 | 536 |
| 24 | Hard | 214.025079 | 86.000000 | 117.453835 | 897 |
| 25 | Hard | 229.465079 | 100.000000 | 99.384014 | 703 |
| 26 | Hard | 230.135011 | 220.000000 | 112.347147 | 859 |
| 27 | Hard | 191.232336 | 106.248001 | 107.666016 | 517 |
| 28 | Hard | 549.760000 | 108.000000 | 107.666016 | 2093 |
| 29 | Hard | 218.600952 | 130.000000 | 129.199219 | 833 |

| | diff | duration | bpm_info | bpm_librosa | num_el |
|---|---|---|---|---|---|
| 30 | Expert | 156.639819 | 164.000000 | 107.666016 | 621 |
| 31 | Expert | 192.000000 | 200.000000 | 99.384014 | 594 |
| 32 | Expert | 164.193741 | 152.000000 | 151.999081 | 459 |
| 33 | Expert | 121.203084 | 91.500000 | 92.285156 | 322 |
| 34 | Expert | 217.646168 | 130.106003 | 129.199219 | 600 |
| 35 | Expert | 155.421315 | 105.000000 | 143.554688 | 703 |
| 36 | Expert | 201.314286 | 178.000000 | 117.453835 | 961 |
| 37 | Expert | 278.502948 | 125.000000 | 129.199219 | 755 |
| 38 | Expert | 317.504989 | 125.000000 | 123.046875 | 1258 |
| 39 | Expert | 264.639048 | 124.000000 | 123.046875 | 1392 |

Figures 14-15: Selected data for Hard and Expert difficulties

| | diff | duration | bpm_info | bpm_librosa | num_el |
|---|---|---|---|---|---|
| 40 | ExpertPlus | 339.888254 | 130.0 | 129.199219 | 1721 |
| 41 | ExpertPlus | 298.460544 | 127.0 | 129.199219 | 1229 |
| 42 | ExpertPlus | 249.460862 | 159.0 | 161.499023 | 916 |
| 43 | ExpertPlus | 282.031338 | 200.0 | 99.384014 | 1863 |
| 44 | ExpertPlus | 215.088390 | 135.0 | 135.999178 | 945 |
| 45 | ExpertPlus | 195.000045 | 138.0 | 135.999178 | 952 |
| 46 | ExpertPlus | 238.205079 | 94.0 | 123.046875 | 1287 |
| 47 | ExpertPlus | 197.435828 | 130.0 | 129.199219 | 1135 |
| 48 | ExpertPlus | 299.742041 | 260.0 | 129.199219 | 2237 |
| 49 | ExpertPlus | 186.328118 | 190.0 | 95.703125 | 812 |

Figure 16: Selected data for ExpertPlus difficulty

To figure out dependencies of num_el with other parameters, the next charts were built: num_el / duration on Figure 17, num_el / bpm_librosa on Figure 18, num_el / bpm_info on Figure 19, and num_el / (duration * bpm_info) on Figure 20.

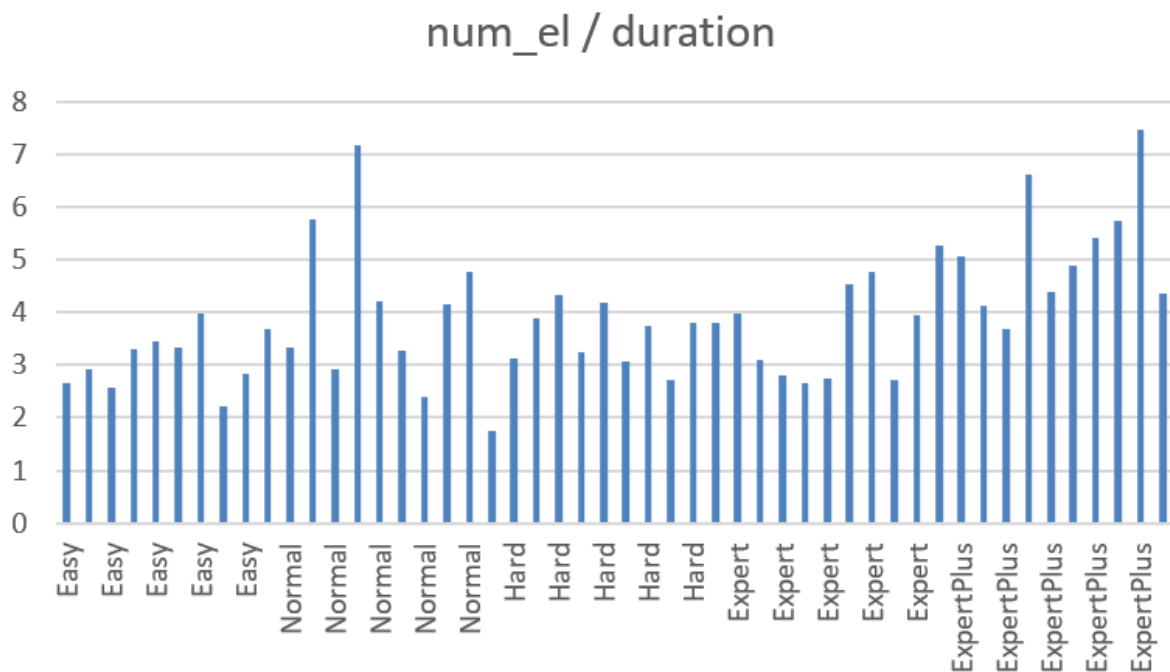

Figure 17: Num_el to duration correlation

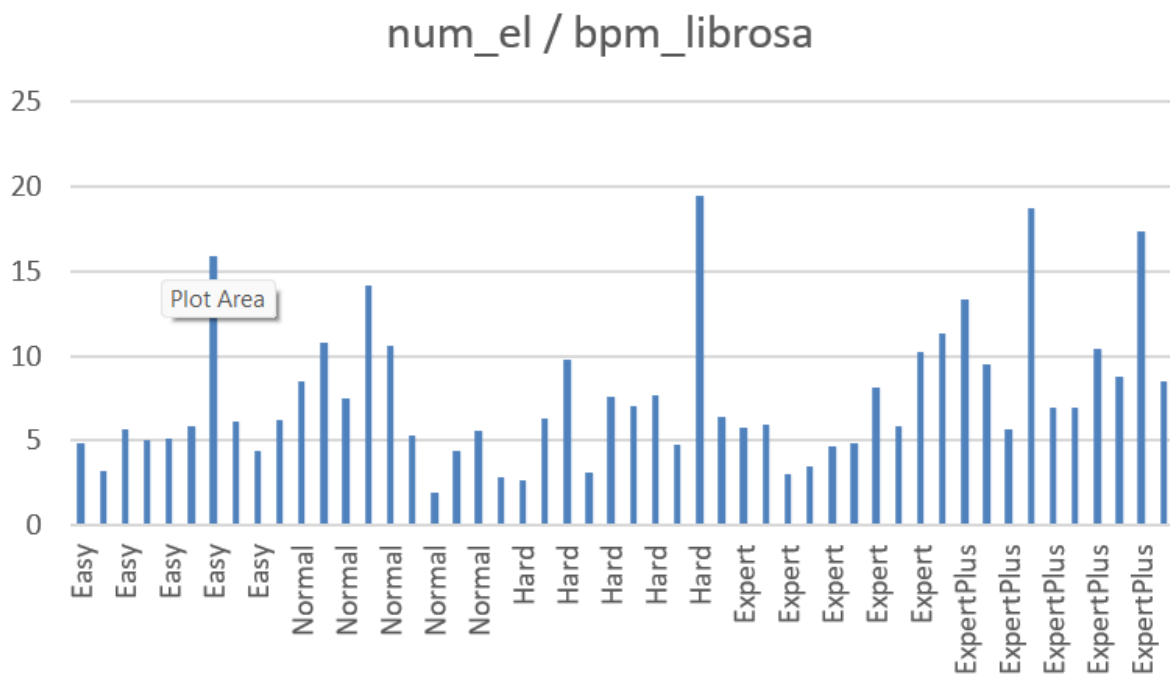

Figure 18: Num_el to bpm_librosa correlation

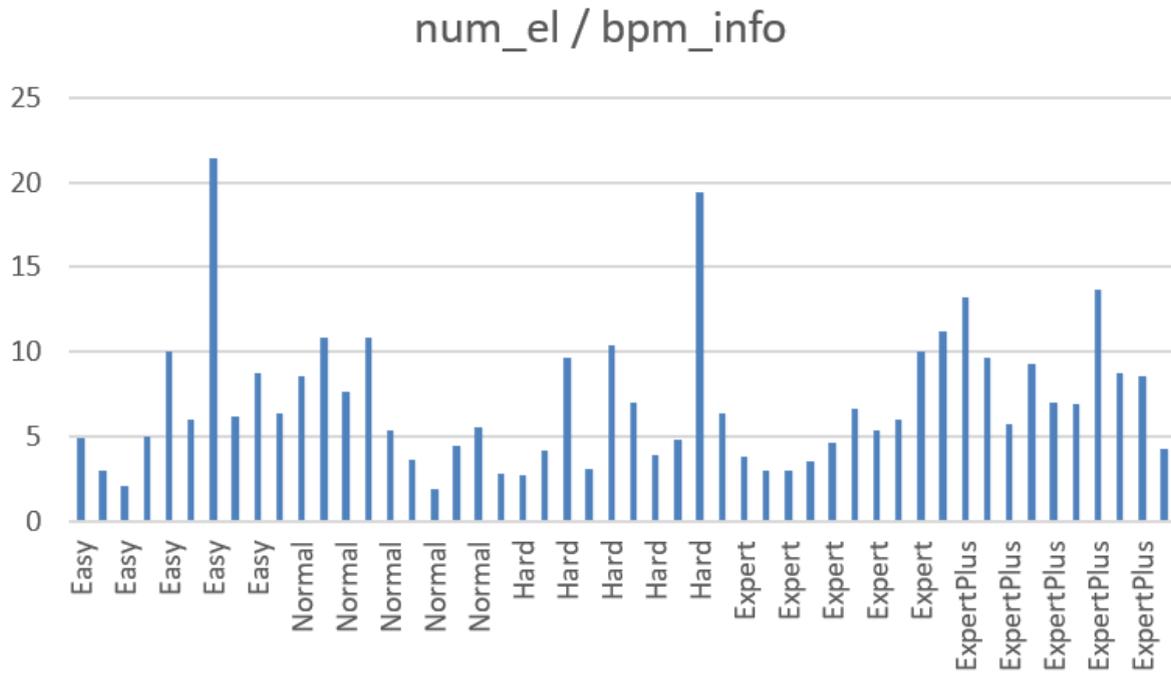

Figure 19: Num_el to bpm_info correlation

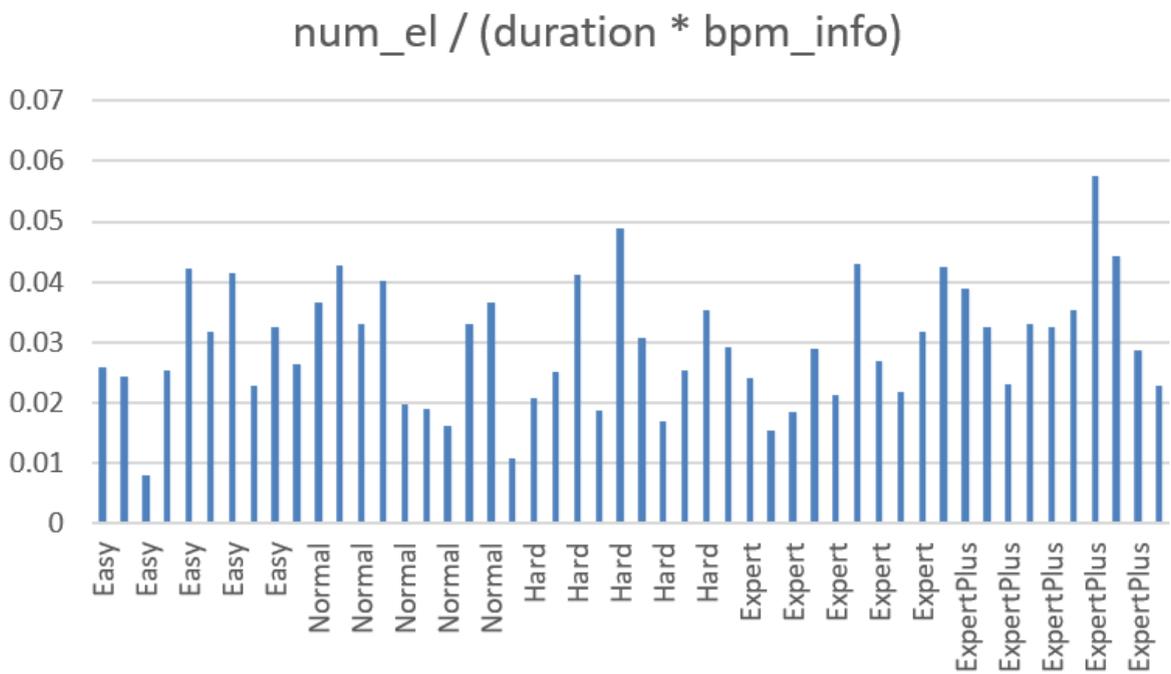

Figure 20: Num_el to duration * bpm_info correlation

It is not obvious from the charts above the numerical difference between different levels, so for each difficulty, minimum, maximum, and average of num_el / duration is calculated. The results are shown in Table 1.

| Difficulty | Min | Max | Average |
|---|---|---|---|
| Easy | 2.223024 | 3.974169 | 3.093952 |
| Normal | 1.745079 | 7.161065 | 3.851733 |
| Hard | 2.703518 | 4.315262 | 3.60789 |
| Expert | 2.656698 | 5.259995 | 3.572549 |
| ExpertPlus | 3.671919 | 7.463084 | 5.071498 |

Table 1: Minimum, maximum, average of num_el / duration

So for Easy levels there should be around 3.09 beats for second, for Normal, Hard, Expert levels - around 3.65 and for ExpertPlus - 5.07. In practice, selecting the maximum amount of peaks (distance = 1) allows for generating a suitable amount of beats for any level of difficulty.

## 4.2 Action selection

For the action selection task, all difficulties were combined as combinations do not become more complicated as the difficulty level increases. The total length of preprocessed data was 915012 elements. In total, there were 15 unique elements, which were saved into an array $all\_characters$. That means at any timestamp, there can be one of these 15 elements that represent 0, 1, 2, or more runes at the same time. The frequency of each possible element is presented in Table 2.

| Element | Count |
|---------|-------|
| n9n9n9n901n9n901n9n9n9n9 | 37389 |
| n9n9n9n9n9n9n901n9n9n9n9 | 147836 |
| n9n9n9n9n9010101n9n9n9n9 | 4 |
| n9n9n9n901n90101n9n9n9n9 | 2 |
| n9n9n9n9n9n90101n9n9n9n9 | 23993 |
| n9n9n9n9n90101n9n9n9n9n9 | 48583 |
| n9n9n9n9n901n901n9n9n9n9 | 27323 |
| n9n9n9n9n9n9n9n9n9n9n9n9 | 28159 |
| n9n9n9n9n9n901n9n9n9n9n9 | 197870 |
| n9n9n9n90101n9n9n9n9n9n9 | 26163 |
| n9n9n9n9n901n9n9n9n9n9n9 | 200806 |
| n9n9n9n901n901n9n9n9n9n9 | 26007 |
| n9n9n9n901010101n9n9n9n9 | 2 |
| n9n9n9n901n9n9n9n9n9n9n9 | 150871 |
| n9n9n9n9010101n9n9n9n9n9 | 4 |

Table 2: Elements frequency in preprocessed data

In this table there are 4 elements that occur in data not more than 4 times. These are outliers, as in-game, it is possible to have a maximum of two runes at the same timestamp. The player has only two hammers, so it is not possible to crash more than two runes simultaneously. So these elements were removed from the data.

The element 'n9n9n9n9n9n9n9n9n9n9n9n9' represents no runes at the timestamp. It is redundant as the element can be removed from _notes and it will have the same result. So this element is removed as well.

The task is dealt as a task of text generation, but instead of 26 letters, 10 unique elements are used. To make inputs out of this big string of data (915012 elements), the data was splitted into chunks with size 200.

This model takes as input the character for step $t_{-1}$ and is expected to output the next character $t$. There are three layers - one linear layer that encodes the input character into an internal state, one GRU layer that operates on that internal state and a hidden state of size 100, and a decoder layer that outputs the probability distribution.

Each chunk is turned into a tensor by looping through the elements of the chunk and looking up the index of each element in $all\_characters$. Finally a pair of input and target tensors for training is assembled, from a random chunk. The input will be all elements up to the last, and the target will be all elements from the first. So if chunk is "123" the input will correspond to "12" while the target is "23".

To evaluate the network one character is fed at a time, the outputs of the network are used as a probability distribution for the next character, and so on. To start generation a priming string is passed to start building up the hidden state, from which then one character is generated at a time.

Then other training parameters are defined:

- Number of epochs: 2000
- Initial learning rate: 0.0001
- Decoder optimizer: Adam optimizer

- Criterion: Cross Entropy Loss
- Scheduler: ExponentialLR with gamma 0.995

You can find plots of loss change during training on Figure 21 and plots of learning rate change on Figure 22.

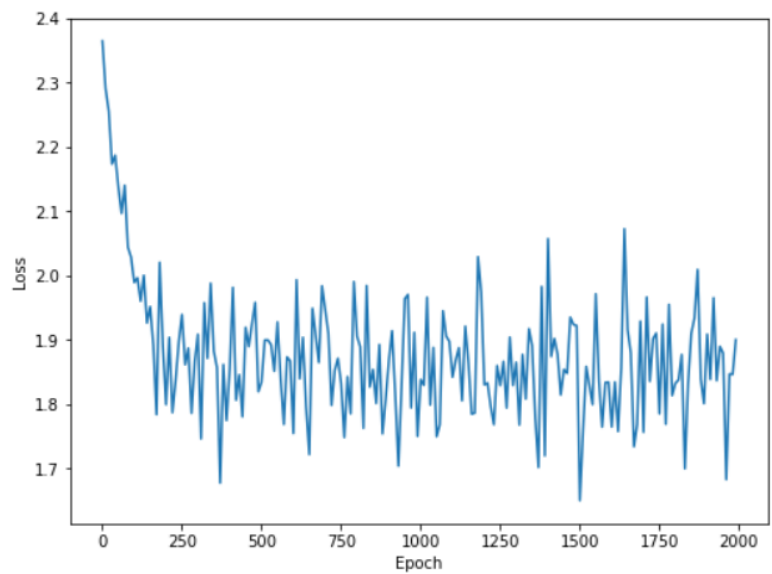

Figure 21: Loss change during training

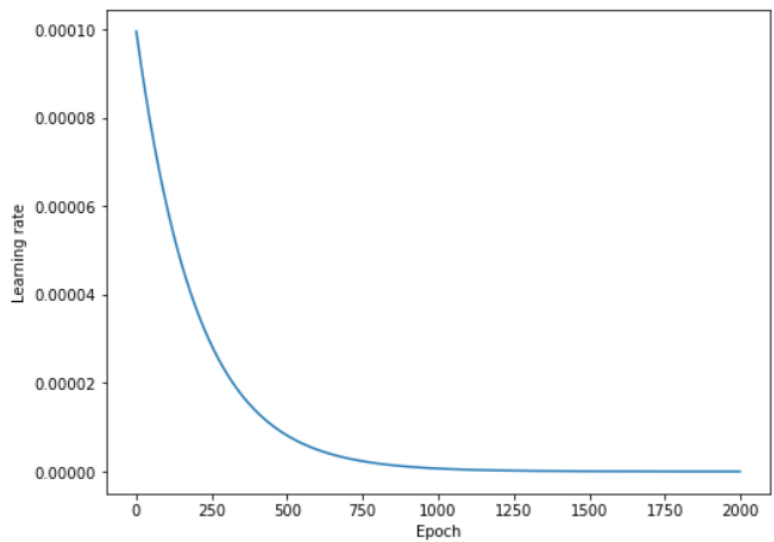

Figure 22: Learning rate change during train

Learning rate is decreased on every iteration as shown on Figure 22. A large learning rate in the beginning allows the model to learn faster and a smaller learning rate in the end may allow the model to learn a more optimal or even globally optimal solution. As shown on Figure 21 there is no significal loss change after 250 epoch, which means the model learnt main patterns.

After training is finished, the evaluate function is used. Every time a prediction is made the outputs are divided by the "temperature" argument passed. Using a higher number makes all actions more equally likely, and thus gives "more random" outputs. Using a lower value (less than 1) makes high probabilities contribute more. As the temperature is turned towards zero only the most likely outputs are being chosen.

For further use in web application, the trained model was saved into "model.pth" file, the unique elements were saved into "all_characters.json" and the parameters (hidden_size and n_layers) were saved into "params.json".

### 4.3 Application

Model itself is not very convenient for common users to use, as it requires certain technical skills. To solve this problem the web application was created. It consists of 2 parts: frontend and backend. The working pipeline is follow:

1. The user goes to a website. You can find a screenshot of this website on Figure 23
2. The user clicks the "Upload File" button and select the song to use
3. The user select different level difficulties. As described on web application the possible level difficulties are:
    - Easy: Level 1-2
    - Normal: Level 3-4

- Hard: Level 5-6

- Expert: Level 7-8

- Expert+: Level 9-10

4. The user clicks the "Download Files" button

5. The user unzip downloaded folder and locate it into the game

Steps 2, 3 and 4 are shown on Figure 24.

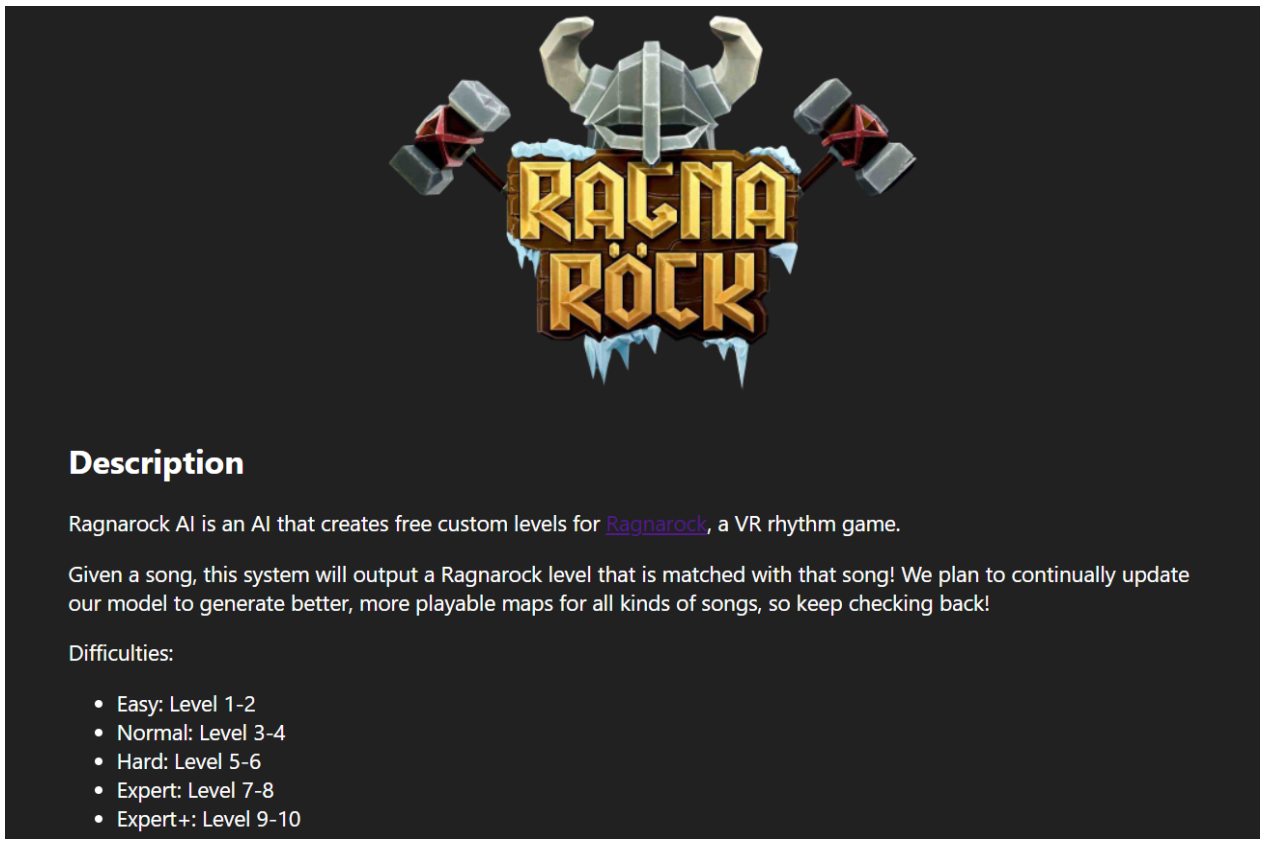

Figure 23: Web application part 1

Figure 24: Web application part 2

To create frontend JavaScript and React were used. After user selection of song and level difficulties, frontend sends request to backend with file of song and other information. The backend receives this data and use evaluation mode of model described above to generate new levels.

The following pipeline for backend is used:

1. The data is received from frontend
2. Using the files "model.pth", "all_characters.json" and "params.json" the trained model is loaded
3. The prime element is selected randomly across all unique elements from "all_characters.json"
4. The prediction length is selected based on level difficulty and song duration, the harder difficulty the more elements there is supposed to be

5. The function evaluate is used with following argument: model, prime element, prediction length and temperature. It returns a list of prediction length of elements

6. The list is decoded back into a JSON format the game can understand

7. The result is sent back to frontend

The backend was built using python and FastAPI. Its pipeline is show on Figure 25.

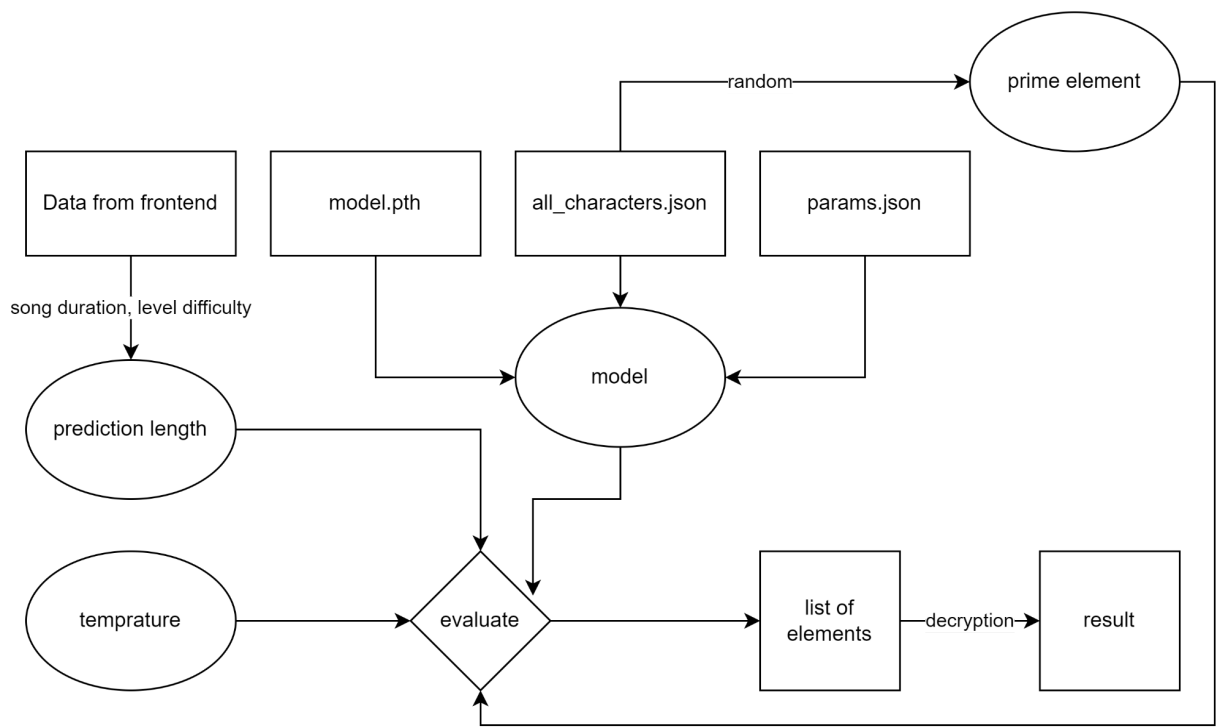

Figure 25: Backend pipeline

# 5 RESULTS

The process of beat map creation, also known as the task of learning to choreograph, for the VR rhythm-based game Ragnarock was automated. The task was broken down into three parts:

1. Action placement: the timing of the beats determination
2. Action selection: determining where in space the runes connected with the chosen beats should be placed
3. Application creation: an end-to-end application for non-technical users.

The data (831 songs) was parsed from the site with custom maps created manually by users to build the dataset. The following information was parsed for each data point: song id, song name, song, author, song file name, bpm, whether easy, normal, hard, expert or expert plus difficulty exists. Everything was saved into a data frame. There were 572 Easy, 130 Normal, 95 Hard, 147 Expert, and 186 Expert Plus levels.

The data was read in the format of dictionaries and then elements in these dictionaries were substituted by unique ids and transformed into a list to process data further for each level file. After that, the data was ready for use in models.

For the first task of action placement, PLP was used to figure out where and how many beats are supposed to be placed. After analyzing the amount of elements distribution for different levels, the maximum amount of peaks for any level difficulty was selected as optimal.

For the action selection task, 915015 elements were used for model training. There were 15 unique elements in the data. 4 of them were removed as an outlier and 1 - as a redundant element. The one-layer GRU with hidden size 100 was trained. The model and other files ("all_characters.json" and "params.json") were saved for further model recreation and use in the web application.

For the last task of application creation, JavaScript with React was used to build the frontend. The user can upload any song file, select different difficulties and download new levels there. The frontend is supported by a backend built with python and FastAPI. The backend receives a song file, and other info in JSON format uses a custom algorithm for action placement, loads a trained model for action selection, save everything in one folder, and sends it back to the user as a zip folder. Everything was hosted online.

# REFERENCES

[1] https://www.ragnarock-vr.com/home

[2] https://beatsaber.com/

[3] DONAHUE, Chris; LIPTON, Zachary C.; MCAULEY, Julian. Dance Dance Convolution [online]. 2017. Available from arXiv: 1703.06891 [cs, stat]

[4] Luc, Ronald. (2020). DeepSaber: Deep Learning for high dimensional choreography.

[5] https://ragnacustoms.com

[6] Grosche, P., & Muller, M. (2011). "Extracting predominant local pulse information from music recordings." IEEE Transactions on Audio, Speech, and Language Processing, 19(6), 1688-1701.

[7] MOORE, Ben. The Best VR Games for 2019 [online] [visited on 2020-02-22]. Available from: https://www.pcmag.com/news/the-best-vr-games-for-2019.

[8] Wu Shiqiang, Huang Ying, Guan Hu, Zhang Shuwu, Liu Jie, "A Spread Spectrum Based Audio Watermarking Method with Embedding Positions Adaption Using Predominant Local Pulse Extraction", 2021 International Conference on Culture-oriented Science & Technology (ICCST), pp.324-329, 2021.

[9] Fikana Mahardika Cantri, Zaqiatud Darojah, Endah Suryawati Ningrum, "Cumulative Scores Based for Real-Time Music Beat Detection System", 2019 International Electronics Symposium (IES), pp.293-298, 2019.

[10] Rumelhart, David E; Hinton, Geoffrey E, and Williams, Ronald J (Sept. 1985). Learning internal representations by error propagation. Tech. rep.

ICS 8504. San Diego, California: Institute for Cognitive Science, University of California.

[11] Jordan, Michael I. (May 1986). Serial order: a parallel distributed processing approach. Tech. rep. ICS 8604. San Diego, California: Institute for Cognitive Science, University of California.

[12] R. Dey and F. M. Salem, "Gate-variants of Gated Recurrent Unit (GRU) neural networks," 2017 IEEE 60th International Midwest Symposium on Circuits and Systems (MWSCAS), 2017, pp. 1597-1600, doi: 10.1109/MWSCAS.2017.8053243.

[13] J. Chung, C. Gulcehre, K. Cho and Y. Bengio, Em-pirical evaluation of gated recurrent neural networks on sequence modeling, 2014.

[14] C. Paseddula and S. V. Gangashetty, "DNN based Acoustic Scene Classification using Score Fusion of MFCC and Inverse MFCC," 2018 IEEE 13th International Conference on Industrial and Information Systems (ICIIS), 2018, pp. 18-21, doi: 10.1109/ICIINFS.2018.8721379.

[15] HOCHREITER, Sepp; SCHMIDHUBER, Jürgen. Long Short-Term Memory. Neural Computation [online]. 1997, vol. 9, no. 8, pp. 1735 1780 [visited on 2020-07-04]. ISSN 0899-7667. Available from DOI: 10.1162/neco.1997.9.8.1735.

**ANNEX A**

The code of the program implementation of the bachelor's qualification work can be found at the link to GitHub:

https://github.com/MeriDK/Thesis